\documentclass[12pt,preprint]{aastex}
\usepackage{amssymb,amsmath,float}
\begin{document}

\title{\large\bf Non-thermal intracluster medium: a simultaneous
interpretation of the central soft X-ray excess and WMAP's detection of 
reduced Sunyaev-Zel'dovich Effect}

\author{Richard Lieu$\,^{1}$ and John Quenby$\,^{2}$}

\affil{\(^{\scriptstyle 1} \){Department of Physics, University of Alabama,
Huntsville, AL 35899.}\\}

\affil{\(^{\scriptstyle 2} \){Blackett Laboratory, Imperial College,
London, SW7~2BZ, U.K.}\\}

\begin{abstract}

WMAP's detection of the
Sunyaev-Zel'dovich effect (SZE) at a much reduced level
among several large samples of rich clusters is
interpreted in terms of conventional physics.  It is by now widely believed
that the central soft X-ray and EUV excess found in some clusters cannot be
of thermal origin, due to problems with rapid gas cooling and
the persistent non-detection of the O VII line, but may
arise from inverse-Compton scattering between
intracluster relativistic electrons and the cosmic microwave background (CMB).
In fact, recently Chandra and XMM 
observations of the soft X-rays from Abell 3112 are well fitted by the sum
of a power law and a thermal virialized
gas component of comparable luminosities. Therefore
the missing SZE flux could simply
be due to an overestimate of the central density of virialized 
electrons which scatter the CMB. 
We also considered if higher energy electrons drawn from the same
power-law population as those responsible
for the soft excess 
may synchrotron radiate in
the intracluster magnetic field of strength B $\lesssim$ 
a few $\mu$G to produce
cluster microwave emissions in the WMAP passbands that account for the
missing SZE flux. Either explanation of the WMAP anomaly
would bolster the current model of central cluster soft excesses, viz.
non-thermal activities prevail in the core of at least some clusters.
The energetic electrons may originate from AGN jet injection, then 
distributed cluster-wide with accompanying {\it in situ} Fermi acceleration,
by Alfven waves. 
However the missing thermal electron explanation provides a less
demanding model.    

\end{abstract}

\section{Introduction}

In a recent paper,
Lieu, Mittaz, Zhang 2006 (LMZ06) published the hitherto most comprehensive 
direct correlation of the {\it Wilkinson Microwave Anisotropy Probe} first year
(WMAP1) data  with the X-ray data of ROSAT and ASCA, in search for
the SZE in the
temperature of the cosmic microwave background
(CMB)  along
the directions to 31 randomly chosen rich clusters located above
the Galactic plane.  The WMAP1 passbands being analyzed cover
the frequency range of 41 -- 94 GHz.  This 
investigation led to the astonishing
finding that on average the
level of SZE detected by WMAP1 is no deeper than the intrinsic CMB
primary anisotropy as seen by WMAP in directions of blank sky (i.e. away
from rich clusters and groups), and in any case
accounts only for 1/3 to 1/4 of the
level expected from the X-ray measurements of the sample clusters, 
Moreover, LMZ06 explored and excluded a variety of possible reasons for
the discrepancy, including emission by radio point sources in the clusters,
which fail by a large margin
to deliver sufficient flux to explain the apparent lack of SZE
in the WMAP1 W-band of 94 GHz.  The results of LMZ06 were corroborated
recently by Bielby \& Shanks 2007, which presented a similar correlation
study between WMAP3 and X-ray observations by Chandra and ROSAT, and likewise
reported a substantially less than expected SZE in the WMAP3 data for a much
larger sample size than the 31 clusters of LMZ06.  The additional
information provided by Bielby \& Shanks 2007 also included a
`truncation test' where the authors abruptly cut off
the X-ray gas profiles of the Chandra cluster sample  at a radius
as small as 2 arcmin, and still found a SZE discrepancy of a factor of
two along the central line-of-sight.  Thus the anomaly is real and
must be resolved.

A separate and superficially unrelated phenomenon of clusters is
the discovery of excess EUV and soft X-ray emission in the 
energy range 0.1 -- 1.0 keV which rises above the
level expected from the spectrum of the hot virialized cluster gas.
This property
has been known for more than one decade to exist in some
clusters (e.g. Lieu et al 1996, Kaastra et al 1999,
Nevalainen et al 2003).  In this paper we demonstrate that it is possible
to connect the non-thermal
inverse Compton interpretation of the cluster soft excess (Hwang 1997,
Ensslin \& Biermann 1998, Sarazin \& Lieu 1998) with the SZE anomaly in
WMAP by extending the dynamic range off this power-law distribution 
of electrons, so that the number density of electrons causing the SZE 
scattering
is actually reduced. 
We also examine the viability of attributing 
any remaining SZE flux discrepancy (after the above effect is
taken into account), to
synchrotron radiation from an unmapped intracluster
population of relativistic electrons. 

Apart from the three 1997-98 papers, there
have been numerous suggestions of a general, non-thermal intracluster 
environment, an early example being
Jaffe (1977) and, more recently,
Quenby et al (1999). 
In particular, a model of the acceleration via Alfven waves driven by major
cluster mergers is given by Brunetti et al (2004).
Of further interest is the idea that relativistic jets carry a significant  
portion of the total energy output of radio galaxies,
causing X-ray emission up to Mpc distance scales 
(Ghiellini \& Celotti 2001).
Celotti, Ghisellini, \& Chiaberge (2001)
provided an analysis of such a scenario for PKS 0637-752.

Unless a separate physical mechanism is proposed (and indeed there is
at least one serious paper on the prospect of neutralino dark matter
decay as a cluster's non-thermal reservoir, see Colafranceso, Profumo, \&
Ullio 2006),
the possibility of cosmic rays 
as the key to solving the soft excess and S-Z puzzles hinges upon
the manner in which energy from AGN is distributed 
widely throughout a cluster, and the speed in which re-acceleration 
can compensate for losses.  It is well known that B\"ohm diffusion happens
too slowly for this purpose (see section 3).
Our model invokes
Alfven waves as the spreading agent which is also
responsible for rapid Fermi statistical acceleration.

\section{Overall cluster non-thermal picture and energy budget}

We assume a continuous non-thermal injection rate into the cluster
environment of
10$^{45}$ ergs~s$^{-1}$
in both electron output and Poynting flux.
This is based upon the analysis of observations by Ghisellini and
Celotti (2001) and the the numbers arising from jet 
injection in the hydrodynamic models of
Zanni et al (2005).  The jet may be  
of cluster size in one dimension (Nulsen et al 2005).
The cluster radius is assumed to
be $\sim$ 1~Mpc and the intracluster magnetic field in the range
1 to 10 $\mu$G. (e.g. Govoni et al 2001;
Medvedev, Silva, \& Kamionkowski 2005).  Ambient gas density is $\sim$
10$^{-4}$ to $10^{-3}$ cm$^{-3}$ (e.g. Brunetti et al 2004).
We also took a typical cluster distance of 400 Mpc, an appropriate
number for the clusters of the LMZ06 sample, since the mean redshift
of the sample is $z =$ 0.1, corresponding to a distance of 428.57 Mpc
in a $\Omega_m =$ 0.3, $\Omega_{\Lambda} =$ 0.7, and $h =$ 0.7
cosmology (Bennett et al 2003, Spergel et al 2006).

While an energetic jet existing a reasonable fraction of the lifetime
of a large, relaxed cluster can transport
energy in one dimension on a cluster scale of 1.0 Mpc
in $10^{9}$ year for 
a jet propagating at 1000 km~s$^{-1}$, a
more general three dimensional dispersion mechanism is necessary
to account for the distributed acceleration required here. 
We adopt a B=6 $\mu$G magnetic
field, tangled on a $\leq 1$ kpc scale over the central
0.5 Mpc region and a density of 10$^{-3}$ cm$^{-3}$ (eg Coma cluster,
Fretti et al., 1995) as representative parameters. 
Alfven waves then move with velocity $ V_{\rm A}=B/\sqrt{4 \pi \rho} \approx$
410 km/s, allowing the transport of 
non-thermal energy over a cluster volume, radius 0.4 Mpc, in
$ 10^{9}$ years.  Instabilities within the jet
and at the edges may produce the Alfven waves 
which continue to spread the energy
after the bulk flow is dissipated.  Rayleigh-Taylor and
Kelvin-Helmholz instabilities in particular were mentioned by Zanni et al
(2005).
These authors show how the disturbance in density and   
entropy could have spread after the jet and shock were switched off, the
computations being followed up to $10^{9}$ years.

Medvedev, Silva, \& Kamionkowski (2005) computed a model of cluster-wide field 
amplification caused by the non-relativistic
Weibel instability, although only 0.1 \% of the total energy ends up in 
the magnetic field.  They claim that the short wavelength
turbulence initially generated gives rise to much longer wavelengths on a 
cosmological time scale. Nishikawa et al, 2003, provide an example of this
instabilty generating turbulence in a relativistic jet.
Merger shocks and intense stellar winds are alternative
sources of heat input to the cluster medium which may 
add to the wave energy. Fretti et al (2004)
developed this idea and fitted synchrotron and inverse Compton 
models to observed spectra, based on giant radio halos.

\section{Cluster soft excess and power law component}

\subsection{Soft excess}

A non-thermal component to the electron spectrum is indicated by 
measurements of a cluster soft excess from regions
where thermal gas at a temperature considerably lower than virial 
is not likely to exist.  
Sarazin and Lieu (1998) estimated the intracluster cosmic ray electron energy 
necessary to account for the observed EUV luminosity of
some clusters as due to
inverse Compton (IC) interaction between the electrons and
the CMB. The IC radiation power is predominantly in the EUV 
and soft X-rays waveband, when the CMB scatters off
lower energy (300 $\lesssim \gamma \lesssim$ 1,000) electrons.
They quoted
\begin{equation}
L_{{\rm IC}}=\frac{4}{3}\frac{\sigma_{\rm T}}{m_{e}c^{2}}
\bar{\gamma}U_{{\rm CMB}}E_{{\rm CR}}
\end{equation}
and demanded
\begin{equation}
E_{{\rm CR}} \sim 2.4\times10^{62}\left(\frac{L_{{\rm EUV}}}
{10^{45}~{\rm ergs~s}^{-1}}\right)
\left(\frac{\bar{\gamma}}{300} \right)^{-1}~~{\rm ergs}
\end{equation}
For A1795 $L_{{\rm EUV}}$ may be as high as 10$^{45}$~ergs~s$^{-1}$
(Mittaz, Lieu, and Lockman 1998), while for the Coma cluster it is $\sim$
a few $\times$ 10$^{42}$ ergs~s$^{-1}$ (Lieu et al 1999).  Thus 
$E_{{\rm CR}}$ ranges from 10$^{59}$ to 10$^{62}$ ergs~s$^{-1}$.
We will take the larger of the 
two values of cosmic ray energy content, viz. $10^{62}$ erg,
as requiring explanation,  Further, we map
the IC output photons to electrons in the energy range just above 100 MeV,
and assume that the electron spectrum cuts off at energies much
below this value.

Recently, the report of strong soft X-ray excess in the cores of
the clusters AS1101 (Werner et al 2007) and A3112 (Bonamente et al 2007)
bolsters the non-thermal interpretation of the soft excess from cluster
centers.  The alternative model which invokes a warm
thermal component to account for the excess must explain how this component
can co-exist with the hot virialized cluster medium without being
clumped into rapdily cooling high density clouds.  Moreover, stringent
upper limits on the amount of intracluster O VII were derived from these
observations, due to the absence of detectable
signatures of O VII emission, the most sensitive tracer of warm gas
hitherto available (see also Lieu \& Mittaz 2005).
The 2-7 keV luminosity of the power-law population
of emitted photons required to fit the soft excess data is an
appreciable fraction of the total X-ray luminosity in this passband,
reaching 30 \% in the case of A3112 (Bonamente et al 2007).

\subsection{Power law and hard excess}
 
To understand how the presence of non-thermal electrons could mean
a significant reduction in the total density of electrons in the
region of concern of a cluster, let us
compare the radiative losses of a single electron by thermal bremstrahlung
and inverse Compton scattering. 
In an electron-proton plasma, the thermal loss at energy $E$
and velocity $\beta c$ is
\begin{equation}
P_{{\rm thermal}} = \frac{dE}{dt}=-\frac{\alpha r_{e}^{2}}{\beta^{2}}cn\bar{g}E
\end{equation} 
for plasma density $n$, Gaunt factor $\bar{g} \approx 1$, 
electron radius $r_{e}$ and
fine structure constant $\alpha$. For inverse Compton losses at electron
Lorentz factor $\gamma$, 
\begin{equation}
P_{{\rm non-thermal}} =
\frac{dE}{dt}=-\frac{32}{9}\pi r_{e}^{2} c n_{ph} h \overline{\nu} \gamma^{2}
\end{equation}
where $n_{ph} h \overline{\nu} $ is the 
CMB energy density. Hence the ratio of thermal to non-thermal power is
\begin{equation}
\frac{P_{{\rm thermal}}}{P_{{\rm non-thermal}}} = \frac{0.67\bar{g}}{\gamma^2}
\end{equation}
at a virialized gas temperature of
kT=9.5 keV.
Hence the required number density of $\gamma=300$ electrons is down by a factor
$\sim 10^{5}$ relative to the number density of 10 keV electrons.

The energetic, cosmic ray electrons in the cluster should not exceed
equipartition with
the magnetic field.   To see that there is no danger here, take
an X-ray luminosity of $5\times10^{44}$ ergs s$^{-1}$
for the central 0.5 Mpc cluster region and, using the IC loss rate  of
\begin{equation}
\frac{dE}{dt}=2.6 \times10^{-14} \gamma^{2}U_{{\rm CMB}} =
1.04 \times 10^{-26}~\gamma^{2}~{\rm ergs}~
{\rm s}^{-1}.
\end{equation}
one obtains a
cosmic ray number density of $2\times10^{-8}$ cm$^{-3}$ and an energy density
$5\times10^{-12}$ ergs cm$^{-3}$; this is to be compared with an equivalent
equipartition field, which is as high as 10$\mu$ G.

Evidence for cross-check the determination of
a cluster's mass by attributing all its X-ray luminosity to a hot
virialized gas came
from weak gravitational
lensing.
Smail et al. (1997) find a $75\%$ efficiency correction is needed to the 
lens shear strengths to bring X-ray determined masses into agreement. A lower
X-ray cluster mass resulting from some of the emitted 
X-rays being of non-thermal origin would imply
a higher efficiency of shear measurement, i.e. the correction factor of
75 \% may not be necessary.
Adoption of the reduced number 
density in the thermal electron cluster component will also lead to an
increase in
the heavy element abundances 
inferred from line intensities because the continuum 
levels are reduced.

\section{The balance between acceleration and loss}

It is not useful to invoke diffusive shock acceleration 
as the distributed cosmic ray electron source since the Larmor radius,
$r_{\rm L} = \gamma m_e c/(eB)$, even at 10$^{13}$ eV
electron energy is only $\sim$ 0.1 pc 
in a 1 $\mu$G field, i.e. diffusive propagation over 
cluster scales is impossible
(see also the second half of this section).  Instead we appeal to
a general Fermi acceleration phenomenon throughout the medium. 
However, the existence of shock accelerated cosmic rays at the boundaries of
the large scale AGN jet driving the additional cluster heating is important for the generation
of the necesaary Alfven wave spectrum.
  
The model assumes a 
continuous injection of energy into at least short wavelength
Alfven waves from the AGN injection, continuing to take place well after switch off of the jet  
as the large scale pressure wave spreads. As the Alfven wave front propagates 
across the cluster, the 
broad spectrum cosmic ray population streams ahead into a
low scattering medium. 
Quasi-linear theory predicts a diffusive scattering 
coefficient (Bell 1978),
\begin{equation}
\kappa_{||}=\frac{4\pi r_{\rm L}v}{3E_{{\rm w}}} \frac{B^{2}}{8\pi}
\end{equation}
where $E_{{\rm w}}(x,p)$ is the energy density of the waves in resonance 
with particles of momentum p
per unit logarithmic bandwidth. 
 Scattering and hence acceleration and wave growth depends on the relation
\begin{equation}
k=\frac{2\pi}{\lambda_{{\rm wave}}}=\frac{v_{\perp}}{v_{||}r_{L}}
\end{equation}
between wave number and Larmor radius.
Bell (1987) shows also that an approximation to quasi-linear theory for the growth rate of
waves due to streaming is
\begin{equation}
\sigma=\frac{4\pi}{3}\frac{V_{A}}{E_{w}}p^{4}v\frac{\partial f}{\partial x}
\end{equation}
$f(p)$ is the particle distribution function.  
The quantity $(4\pi/3)p^{4}vf(p)$
is the pressure per unit logarithmic bandwidth of particles of momentum p.
Assuming equipartition between particle and wave energy at resonance in each energy band
within the turbulent Alfven wave front driven by the region of shock acceleration,
the growth rate in the region originally of lower scattering power is simply
$\sigma=V_{A}/\delta x$ where $\delta x$ is the scale size for the gradient in $f(p)$.
A $\delta x \leq 26$ pc produces $\sigma^{-1}\geq 6\times10^{4}$ years, the time constant 
we will show is needed for acceleration of $2\times 10^{13}$ eV electrons, 
the energy obtained in the most extreme version of the model developed. This may be compared with
a minimum structural size determined by the Larmor radius at this energy when the jet is far out of
$1\times10^{16}$ cm. 

The strong turbulence resulting from the 
streaming instability allows a diffusion mean free path
within an order of magnitude of the B\"ohm 
limit throughout the accelerated particle energy range
(Zank et al 2000). 
This treatment of scattering is in contrast to that 
of Brunetti et. al. (2004) who used
an assumed initial Kolmogorov or 
Kraichman turbulence spectrum determined by a minimum k value
to specify the quasilinear diffusion coefficent 
and then allow the wave particle interactions 
to cause decay of the wave's spectral intensity.

The Fermi acceleration time constant is
\begin{equation}
\tau_{\rm F} = \frac{\lambda v}{3V_{\rm A}^{2}},
\end{equation}
where $v$ is the particle velocity, and $\lambda$ is the diffusion
mean free path.
Experience in measuring $\lambda$ in the turbulent interplanetary medium 
suggests $\lambda\sim 30r_{\rm L}$, rather than the B\"ohm value.

The lifetime of an electron of energy $\gamma m_e c^2$ 
against inverse Compton interactions (losses) 
on the cosmic microwave background (CMB) of energy density 
$U_{{\rm CMB}}$, is
\begin{equation}
\tau_{{\rm IC}}=\frac{\gamma m_{e}c^{2}}
{\frac{4}{3} \sigma_{\rm T} c \gamma^{2} U_{{\rm CMB}}}=
2.31 \times 10^{12} \gamma^{-1}~{\rm years},~~{\rm for}~\gamma \gg 1.
\end{equation}
By equating inverse Compton and Fermi time constants after applying
our adopted parameters, we obtain a maximum permitted 
electron energy of 1.9 $\times$ 10$^{13}$ eV.
and a life of $6.1\times 10^{4}$ years. 
This limit is important to our ensuing calculation of a possible
upper limit to the electron spectrum. With negligible reacceleration,
the loss time at $\gamma=300$ is $7.7\times10^{9}$ years.  

V\"olk, Aharonian and Breitschwerdt (1996) 
reviewed the non-thermal energy content of galaxies and discussed
the contribution of shocks to the cosmic ray content. 
We may consider two alternatives to the Alfvenic heating
model for the supply of high energy electrons. 
The first employs shocks at collisions between clusters and hence parameters
for the medium inside clusters, so $V_{{\rm sh}}=10^{7}$ cm/s and
$B=6\times10^{-6}$~G. If the model to explain the SZE anomaly
requires distributed synchrotron radiation at 41 GHz, the electron $\gamma$
given by $\nu=4.3\times10^{6} \gamma^{2} B$ 
needs to reach $\gamma=4.0\times10^{4}$, where the
IC loss time is $5.8\times10^{7}$ years. 
 Now using as the approximate shock acceleration time  
\begin{equation}
\tau_{{\rm ac}}=\frac{3\kappa_{||}}{V_{{\rm sh}}^{2}}
\end{equation}
where
\begin{equation}
\kappa_{||} = \frac{1}{3} \lambda v
\end{equation}
is the diffusion coefficient, one finds that $\tau_{{\rm ac}} 
\approx 10^{4}$ years.
However, the time to diffuse over a cluster scale of  $R=400$ kpc,
\begin{equation}
\tau_{{\rm diffusion}}\sim \frac{R^{2}}{\kappa_{||}}
\end{equation}
is $10^{15}$ years.
The second alternative concerns relativistic AGN jets 
near their source.   We take $V_{{\rm sh}}\sim c$ and $B=6\times10^{-6}$ G.
These parameters yield $\tau_{{\rm ac}} \sim 10^{-3}$  year. The timescales
relating to shock acceleration
indicate that there should be no lack of localised sources of relativistic
electrons, especially in an AGN environment where a relativistic jet
of Lorentz factor $\Gamma$ produces the $\Gamma^{2}$ factor
increase of acceleration first noted
Quenby and Lieu (1989). However, the long diffusion timescale spells the
impossible task of filling the entire cluster with a
non-thermal population
using isolated diffusive shock acceleration sources as input, i.e.
the energetic
particles so produced are expected to emit
synchrotron radiation only locally.
V\"olk et al had similarly found 
extremely long times for the escape of particles from clusters.      
  
\section{Possible Cluster 50 GHz Emission}

\subsection{Synchrotron Emission}

As an alternative, or even an addition to the model explaning of the SZE, the
extension of the cluster non-themal population to energies sufficient to provide
synchrotron radiation in the WAMP response range is expolored. 
Without a detailed, non-linear model of the cosmic ray cluster electron spectrum,
which is beyond the scope of this work, we cannot predict the overall spectral shape.
Models without particle-wave back reaction and relying only on synchrotron loss produce
a pile up at high energies before a cutoff (Borovsky and Eilek, 1986). A typical model where
turbulent driving of waves occurs, in this case from eddies inherent in jet turbulence radiating
waves, is studied by Eilek and Henriksen (1984). They find power a power law
 electron and synchrotron spectrum can arise. Here we will assume a power law,
 but only need it to relate low energy electrons yielding Inverse Compton photons
and high energy electrons yielding synchrotron photons without need of detailed 
knowledge of the shape in between.
Take the intracluster cosmic ray electron spectrum as
\begin{equation}
 \frac{dn(\gamma)}{d\gamma} =\frac{N}{4\pi} \gamma^{-s}~~{\rm cm}^{-3}~
{\rm sr}^{-1}.
\end{equation}
The power output is
\begin{equation}
\frac{d^2 P}{dV d\nu}=1.7\times10^{-21}NB^{\frac{s+1}{2}} \left(
\frac{4.3 \times 10^6}{\nu} \right)^{\frac{s-1}{2}}~~{\rm ergs}~{\rm cm}^{-3}~
{\rm Hz}^{-1}.
\end{equation}
Now the peak frequency of the synchrotron spectrum is given by
\begin{equation}
\nu \approx 4\times10^{6} \gamma^2 B~~{\rm Hz}
\end{equation}
If the lower limit to $\gamma$ could correspond to $\gamma \approx$ 200, or
electron energy $\approx$ 100 MeV, below the peak of the cosmic ray electron
flux in our Galaxy, the minimum emitted frequency is
$\nu_{{\rm min}}=$ 1.7 $\times$ 10$^{5}$ Hz.  
The upper emitted frequency, corresponding to the electron cutoff energy of 
1.9 $\times$ 10$^{13}$ eV as explained after Eq. (11), is
$\nu_{{\rm max}}=$ 3.4 $\times$ 10$^{16}$ Hz. The cluster volume assumed is
$V_{{\rm vol}}=1.4\times10^{73}$ cm$^{3}$.
To explain the anomalies in the S-Z observations, a synchrotron power
output $\approx$ 20,000 times less than that
of the CMB  (i.e. $\approx 10^{-19}$ 
ergs cm$^{-2}$ s$^{-1}$ sr$^{-1}$ Hz$^{-1}$) is required at 41 GHz.
Thus the power at
this frequency is defined for equation (16) and represents
the measurement at one energy of
the assumed cosmic ray electron spectrum.   

The second measurement the electron specrum 
is afforded by the cluster soft excess.  
More precisely, by taking $10^{62}$ ergs as the
cosmic ray energy content and linking
the IC photons to the electron energy range just above 100 MeV  for an
electron spectrum that cuts off below this value, a second point
on the electron spectrum is defined.

\subsection{Electron spectrum required by the SZE anomaly and soft excess}

The electron spectrum may be recast into a
more recognizable form, as
\begin{equation}
n(\gamma)d\gamma=\rho(E)dE=\rho_0 E^{-s}dE
\end{equation}
where
\begin{equation}
\rho_0 =(m_{e}c^{2})^{s-1} \frac{N}{4\pi}.
\end{equation}
A typical set of spectral parameters satisfying both the SZE anomaly and the 
EUV excess are:
$s=5.92, 
N=4.13\times10^{5}$ cm$^{-3}$ sr$^{-1}$ and 
$\rho_0=$4.0 $\times$ 10$^{32}$ cm$^{-3}$ eV$^{s-1}$, 
applicable to the energies
between 100 MeV and 2 $\times$ 10$^{4}$ GeV, with an ensuing
intracluster relativistic electron number
density of  $\sim$ 2.0 $\times$ 10$^{-8}$ cm$^{-3}$.
For comparison, our Galaxy 
contains a cosmic ray proton number density 
of 6 $\times$ 10$^{-8}$ cm$^{-3}$.
There is thus {\it no difference}
between the requirements on acceleration in the intracluster medium
containing a central AGN and the cosmic ray flux in a normal galaxy. 

To ensure that the model works self-consistently, we emphasize that
the proposed intracluster non-thermal electrons will not by themselves 
produce any obvious SZE signal 
above the normal level from the intracluster
hot gas.
Moreover, their
synchrotron radiation at radio frequencies
is below normal radio astronomy sensitivity limits.  A typical
cluster observation has its lowest contour at 1~mJy over a 43 arcsec beam 
resolution. This corresponds to a background
from the cluster of 10$^{-17}$~ergs~cm$^{-2}$~s$^{-1}$~sr$^{-1}$~Hz$^{-1}$.
Thus, unless we contemplate a larger  cosmic ray population than
depicted by Eq. (16) using maximum associated parameters,
it would require an exceptionally low noise observation to pick out the aspect
of cosmic ray signal suggested here.
There are indeed some large clusters where a 
radio halo appears. Giovannini et al (1999)
found $\sim$ 5 \% of a complete X-ray selected cluster sample to have 
diffuse radio emission.  Synchrotron radial profiles more extended
than that of X-rays can produce 2 -- 2.5 Mpc source
sizes, while mini-halos confined to central regions and radio relics
on the cluster periphery were also reported (Brunetti 2004).

\subsection{Gamma-Ray Limits}

There are some indications of non-thermal, hard X-ray cluster emission, 
but a typical EGRET gamma ray upper
limit (on the Coma cluster, Sreekumar et al,
1996) is 3~$\times$~10$^{-8}$~photon~cm$^{-2}$~s$^{-1}$ at 100 MeV. 
The higher $\gamma$ end of our proposed
electron power-law will produce gamma rays, also by the IC effect on the CMB. 
For a particular energy of the emerging photon the required electron 
$\gamma$ is given by
\begin{equation}
 h\nu_{{\rm IC}}=\frac{4}{3}\gamma^{2} h\nu_{{\rm CMB}},
\end{equation} 
and using the equation for the rate of IC energy loss for an electron,
 we find an electron density $4\times10^{-24}$
electrons cm$^{-3}$  above the recoil threshold for IC induced 100 MeV gamma 
emission. Assuming most of the emission is concentrated at
around the threshold of $\gamma\sim$~6~$\times$~10$^{5}$,
and a cluster distance of 100 Mpc, 
the flux at earth is $\sim$ 1~$\times$~10$^{-11}$ photon~cm$^{-2}$~s$^{-1}$, 
i.e. even under the
scenario of maximal cosmic ray pressure the EGRET limit is not
violated by our model.
  
\section{Summary and conclusions}

Although the basic ideas of utilizing AGN activity and Alfvenic heating 
as a prime energy source for
the intracluster medium have been proposed, we focussed upon 
a number of specific non-thermal cluster processes.
Our model started with the limited spatial extent of the AGN related jet 
feeding.  Alfven waves then
distribute the energy within a cluster time frame, resulting in a cluster-wide
population of relativistic electrons with a power-law spectrum.
The chief constraint on electron lifetime, and hence spectral hardness,
is inverse Compton loss on the CMB. 
Provided a reasonable fraction of the observed cluster AGN power can
spread outwards, there will be enough energy to explain the WMAP1 SZE 
anomaly of LMZ06 and the cluster soft excess phenomenon.
The proposed mechanism is in-line
with previous ideas in interpreting giant radio halos, although we treat the 
generation of the intracluster wave spectrum in a different manner.

Among the two models offered, the simplest explanation 
of the WMAP SZE anomaly is that
the cluster electron population has hitherto been overestimated,
because all the emitted X-rays 
from a cluster were attributed to the virialized hot thermal medium when
as many as half of these thermal electrons should in fact be replaced by
a much smaller population (in terms of number density) of
cosmic ray electrons accelerated to
modest Lorentz factors.
After correcting for this effect, the 
result is a significant reduction of the predicted SZE.   
In the alternative (more indirect, and hence perhaps
less likely) approach to solving the observational
problems,  we invoked synchrotron
radiation in the microwave frequency range from the higher energy end of
the same power-law distribution of cluster cosmic ray
electrons as that responsible for the SZE anomaly.  Such electrons can exist
only if a continuous input from Alfvenic acceleration at high Lorentz
factors (to combat IC losses) is available. 

The salient features of  our proposed
model to reduce the thermal electron content
of clusters are summarized as follows.
(a) For non-thermally active clusters the WMAP SZE anomaly would
not implicate negatively upon the cosmological
origin of the CMB.  Rather, it probes the properties
of the intracluster medium, which (for some 
clusters at least) harbor a cosmic ray energy
density approaching that of our own Galaxy.
(b) Contribution to cluster mass from the thermal medium are significantly 
overestimated in clusters with 
anomalous SZE. Cosmic abundances may be underestimated.
(c) Anomalous SZE could arise from clusters with one or more of the following
special characteristics: powerful AGNs,
cluster scale X-ray jets, long radio jets,
radio `ridges', radio halos, or other evidence for recent cluster merger.
(d) Similarly, soft 
excess emissions are more likely where there is evidence of large-scale
cluster turbulence, as above.
(e) Extended cluster radio emission identified in some massive clusters
could be explained by
an enhanced version of the same Alfvenic heating and IC loss
balance presented herein.
(f) Relatively higher fluxes of non-thermal X-rays 
are emitted by clusters with anomalous SZE.
(g) The full SZE is expected to be found in relaxed clusters that
do not exhibit any of the characteristics outlined in section 3.


\newpage 

\noindent
{\bf References}

\noindent
Bell, A R., MNRAS, 1987, 182, 147.

\noindent
Bennett, C.L. et al 2003, ApJ, 148, 1.

\noindent
Bielby, R.M., \& Shanks, T. 2007, MNRAS submitted (astro-ph/0703470).

\noindent
Bonamente, M., Nevalainen, J., \& Lieu, R. 2007, ApJ submitted.

\noindent
Borovsky, J. E., \& Eilek, J. E., 1986, ApJ, 308, 929.

\noindent
Brunetti, G., Blasi, P., Cassano, R., \& Gabici, S., 2004, MNRAS, 350, 1174.

\noindent
Brunetti, G., 2004, IAU colloquium 195~`Outskirts of galaxy clusters: intense\\ 
\indent life in the suburbs', Turin, Italy (astro-ph/0404507).

\noindent
Celotti, A., Ghisellini, G. \& Chiaberge M., 2001, MNRAS, 321, L1.

\noindent
Colafrancesco, S., Profumo, P., \& Ullio, P., 2006, A \& A in press 
(astro-ph/0507575).

\noindent
Eilek, J, A., \& Henriksen, R., N., 1984, ApJ., 277, 820.
 
\noindent
Ensslin, T. \& Biermann, P.L. 1998, A \& A 330, 90.

\noindent
Feretti, L., Orru, E., Brunetti, G., Giovannini, G., Kassim, N., \& Setti, G.\\
\indent 2004, A \& A, 423, 111.

\noindent
Ghisellini, G. \& Celotti, A. 2001, in `Issues of unifications of AGNs', \\
\indent eds. Maiolino, R., Marconi, A. \& Nagar, N. (astro-ph/0108110).

\noindent
Giovannini, G.; Tordi, M.; Feretti, L., 1999, NewA, 4, 141.

\noindent
Govoni, F., Taylor, G. B., Dallacasa, D., Fretti, L. \& Giovannini, G., 2001,
A \& A, 379, 807.

\noindent
Hwang, C. -Y. 1997, Science, 278, 1917.

\noindent
Jaffe, W.J. 1977, ApJ, 212, 1.

\noindent
Kaastra, J.S., Lieu, R, Mittaz, J.P.D., Bleeker, J.A.M., Mewe, R. \& \\
\indent Colafrancesco, S., 1999, ApJ, 519, L119.

\noindent
Lieu, R., Mittaz, J.P.D., \& Zhang, S.N., 2006, ApJ, 648, 176.

\noindent
Lieu, R., \& Mittaz, J.P.D. 2005, in `The identification of dark matter', \\
\indent Ed. N.J.C. Spooner \& V. Kurdryavtsev, p. 18 (astro-ph/0501007).

\noindent
Lieu, R., Mittaz, J.P.D., Bowyer, S., Breen, J.O., Lockman, F.J., \\
\indent Murphy, E.M. \& Hwang, C. -Y. 1996, Science, 274, 1335.

\noindent
Lieu, R., Ip, W. -H., Axford, W. -I., \& Bonamente, M., 1999, ApJ, 510, L25.\\

\noindent
Medvedev, M.V., Silva, L.O., \& Kamionkowski, M., 2005, astro-ph/0512079.

\noindent
Mittaz, J.P.D., Lieu, R., \& Lockman, F.J., 1998, ApJ, 498, L17.


\noindent
Nevalainen, J., Lieu, R., Bonamente, M., and Lumb, D.,
2003, ApJ, 584, 716.

\noindent
Nishikawa, K.-I, Hardee, P., Richardson, G., Preece, R., Sol, H., \& 
Fishman, G.J.,\\
\indent 2003, ApJ, 595, 555.

\noindent
Nulsen, P.E.J., McNamara, B.R.,  Wise, M.W., \& David, L.P. 2005,\\
\indent ApJ, 628, 629.

\noindent
Quenby J. J., \& Lieu, R., 1989. Nature, 342, 654
 
\noindent
Quenby, J.J. et al 1999, Proc. 2nd int. workshop on the identification of\\
\indent dark matter, World Scientific, p137.
 
\noindent
Sarazin C.L. \& Lieu, R. 1998, ApJ, 494, L177.

\noindent
Smail,I. , Ellis, R. S., Dressler, A., Couch, W. J., Oemler, A., Jr., Sharples, R. M., \\
\indent \& Butcher, H., 1997, Ap. J., 479, 70.  

\noindent
Spergel, D. et al, 2006, ApJ in press (astro-ph/0603449).

\noindent 
Sreekumar, P., Bertsch, D. L., Dingus, B. l., Esposito, J. A.,
Fichtel, C. E., \\
\indent Fierro, J., Hartman, R. C., Hunter, S. d., Kanbach, G., Kniffen,
D. A., Lin, Y. C., \\
\indent Mayer-Hasselwander, H. A., mattox, J. R., Michelson, P. F., 
von Montigny, C.,\\
\indent Mukherjee, R., 
Nolan, P. L., Schneid, E., Thompson, D. J., and Willis, T. D., 1996, \\
\indent ApJ. 464, 628.
  
\noindent
V\"olk, H.J., Aharonian, F.A., \& Breitschwerdt, D., 1996, Space Sci. Rev.,
75, 279.

\noindent
Werner, N., Kaastra, J.S., Takei, Y., Lieu, R., Vink, J., \& Tamura, T. 2007,\\
\indent A \& A in press (arXiv:0704.0475).

\noindent
Zank, G. P., Rice, W. K. M., \& Wu, C. C. J. 2000, Geophys. Res., 105, 25079. 

\noindent
Zanni, C., Murante, G., Bodo, G., Massaglia, S., Rossi, P. \& Ferrari, A. \\
\indent 2005, A \& A 429, 399.

\end{document}